\documentclass[prb,twocolumn,superscriptaddress,showpacs]{revtex4}
\usepackage{graphicx}
\newcommand{\beq}{\begin{equation}}
\newcommand{\eeq}{\end{equation}} 
\newcommand{\beqa}{\begin{eqnarray}}
\newcommand{\eeqa}{\end{eqnarray}}
\newcommand{\ba}{\begin{array}}
\newcommand{\ea}{\end{array}}
\begin{document}
\bibliographystyle{prbrev}

\title{Vortex arrays in a rotating superfluid $^4$He nanocylinder}

\author{Francesco Ancilotto}
\affiliation{Dipartimento di Fisica e Astronomia ``Galileo Galilei''
and CNISM, Universit\`a di Padova, via Marzolo 8, 35122 Padova, Italy}
\affiliation{ CNR-IOM Democritos, via Bonomea, 265 - 34136 Trieste, Italy }

\author{Mart\'{\i} Pi}
\affiliation{Departament ECM, Facultat de F\'{\i}sica,
and IN$^2$UB,
Universitat de Barcelona. Diagonal 645,
08028 Barcelona, Spain}

\author{Manuel Barranco}
\affiliation{Departament ECM, Facultat de F\'{\i}sica,
and IN$^2$UB,
Universitat de Barcelona. Diagonal 645,
08028 Barcelona, Spain}

\begin{abstract}
Within Density Functional theory, we investigate
stationary many-vortex structures 
in a rotating $^4$He nanocylinder at zero temperature.
We compute the stability diagram
and compare our results with the classical model
of vortical lines in an inviscid and incompressible fluid.
Scaling the results to millimeter-size buckets, 
they can be compared with 
experiments on vortex arrays conducted in the past.  
Motivated by recent experiments that have used atomic impurities as a means of
visualizing vortices in superfluid $^4$He droplets, we have
also considered the formation of chains of xenon atoms along a vortex line 
and the interaction between xenon
atoms inside the same vortex and on different neighboring 
vortex lines. 

\end{abstract}
\date{\today}
\maketitle

\section{Introduction}

At temperatures low enough, $^4$He droplets and 
confined clouds of ultra-cold boson gases 
are paradigms of superfluid quantum droplets. Small para-hydrogen clusters
are likely superfluid,\cite{Gre00} although no definite conclusion has 
been experimentally drawn yet.\cite{Zen14}
Together with the frictionless displacement of 
impurities at velocities below 
the Landau critical velocity,\cite{Pit03} 
the appearance of quantized vortices is the
recognized hallmark of superfluidity
in liquid $^4$He \cite{Fey55,Vin61} 
that appear at temperatures below 2.17 K,  
the superfluid transition temperature.

Due to its superfluid character, $^4$He remains at rest when its container 
rotates until a critical angular velocity
is reached, leading to the appearance of vortices with 
quantized velocity circulation in units of $h/M$,
where $h$ is the Planck constant and $M$ is the $^4$He  atomic mass.
Free  --or attached to impurities-- linear vortices  
have been theoretically studied  using methods of different complexity, 
see Refs. \onlinecite{Dal92,Ort95,Gio96,Pi07,Gal14,Sad97} and references therein.
In the case of cold gases, vortices
have been nucleated using methods such as
the rotation of the magnetic trap or the thermal cloud
during the evaporative cooling process.\cite{Pit03,Fet09} 
These
methods bear some similarity with the ``rotating bucket'' way of
nucleating vortices in superfluid liquid helium.\cite{And46b}

While vortex arrays in cold gases can be optically identified after the
condensate has expanded upon removing the magnetic or optical trap,
vortex distributions in liquid helium have been only visualized by doping them.
They were first imaged
by Packard and co-workers\cite{Wil74,Yar79} with the spots
of light on a phosphorescent screen caused by the hitting of electrons 
originally attached to vortex lines. More recently, quantized vortices 
have been visualized by suspending micron-sized solid particles
of hydrogen in superfluid  $^4$He at relatively high 
temperatures $T \sim 2$ K,\cite{Zha05,Bew06} where they 
are found to arrange themselves with nearly equal spacing 
along vortex lines, or at lower  temperatures $T < 0.6$ K by He$^*_2$ 
through excimers created {\it in situ} by ionization
in a strong laser field.\cite{Zme13}  Coalescence of gold nano-clusters 
inside vortices in superfluid $^4$He has been observed\cite{Mor10} and further discussed
in Ref. \onlinecite{Gor12}.

The equilibrium configurations of vortex arrays in  
rotating superfluid helium were computed in Refs. \onlinecite{Hes67,Sta68,Cam79}
within the classical vortex theory of an inviscid and incompressible fluid that 
incorporates the quantum effect of quantization of circulation around vortex lines. 
A comprehensive review of the activity on quantized
vortices in superfluid liquid helium before the 1990s can be found
in the book by Donnelly.\cite{Don91}

With the advent of helium droplet experimental facilities in the 
1990s,\cite{Toe04} the issue whether 
nanodroplets are superfluid or not became a subject of intense experimental and 
theoretical activity.\cite{Sin89,Ram90,Chi92,Kro01} Helium droplets are
created by expanding a cold helium gas and attain a limiting 
temperature below 0.4 K,\cite{Bri90,Har95} lower than the superfluid 
transition temperature. The experimental confirmation of superfluidity 
in helium droplets was provided by Toennies and coworkers, who 
established that an OCS molecule inside a $^4$He droplet displays a 
neat ro-vibrational spectrum, indicating that the molecule may rotate 
without dissipation,
at variance with its behavior in a normal-fluid $^3$He droplet.\cite{Gre98}   
It is worth mentioning that the minimum number 
of atoms in the droplet for displaying 
superfluid features is amazingly small, about 60 atoms.

Several theoretical studies have been conducted for a single linear vortex
in helium droplets taking for granted that they
could be nucleated inside them.\cite{Bau95,Dal00,Anc03,Leh03,Sol07}
Experimentally, the appearance of quantum
vortices\cite{Gom12,Spe14,Lat14,Tha14} and the
frictionless displacement of swift impurities in helium 
droplets\cite{Bra13,Mat13} have been recently established. 
In both cases, foreign atoms were used as tracers or swift impurities.
The motion of tracer particles in superfluid $^4$He has been addressed
in a number of papers, see for instance Refs. \onlinecite{Poo05,Bar07,Fie12} 
and references therein.

Quantized ring vortices have been theoretically predicted to accompany the 
sinking of cations produced by photoionization of the neutral species sitting 
at the surface of $^4$He droplets under very well 
controlled experimental conditions.\cite{Mat14} 
Their effect on physical observables that might allow to 
experimentally detect them remains inconclusive yet. 

Very recently, in a femtosecond X-ray coherent diffractive imaging
experiment the existence of vortex arrays has been demonstrated for
helium droplets.\cite{gom14} The diffraction images revealed characteristic
Bragg patterns from Xe clusters trapped in the vortex cores present
in the helium droplets made of $N= 10^7 - 10^{10}$ helium atoms
produced by fragmentation of a cryogenic fluid. 

Theoretically addressing vortex arrays in helium droplets is a
challenge irrespective of the
method one uses. It is currently beyond the 
capabilities of quantum Monte Carlo
methods that even for one single vortex yield different results
depending on whether the fixed node or the fixed phase approximation is
used.\cite{Ort95,Gio96,Gal14} 
To determine the equilibrium configuration of a vortex array
within the classical vortex theory of an inviscid and incompressible fluid, 
it has to be imposed as a boundary condition that 
the vortex lines perpendicularly hit the droplet surface,
which is not a trivial issue.\cite{Bau95,Leh03} This condition is built-in within
the density functional theory (DFT) approach,\cite{Anc03,Pi07} that however has as a 
practical limitation the computing time needed to determine
the structure of large enough droplets capable to host many vortex lines, thus
hampering any systematic study of their
appearance as a function of the rotating angular velocity.

As a first step towards a DFT description of vortex arrays in helium
droplets we present here the simpler case of vortex arrays in a rotating  
self-bound $^4$He nanocylinder infinitely extended along the axial direction. On the
one hand, it will
allow to assess the applicability of the DFT method to vortex array
configurations and on the other hand 
to address the cylindric configuration attained in 
rotating bucket experiments, for whose description
only the classical vortex theory of an inviscid and incompressible fluid 
has been used in the past.\cite{Hes67,Sta68,Cam79}
We complement this study determining for some cases of study, the
structure of vortices doped with Xe atoms because of their
relevance to recent experimental studies of vortex arrays
in superfluid $^4$He nanodroplets.\cite{Gom14}

\section{Model}

We consider a self-bound superfluid $^4$He cylinder rotating
around its symmetry $z$-axis with a constant 
angular velocity $\omega $. We assume in our calculation
a uniform density along the $z$-direction, which implies
that the resulting vortices always remain rectilinear and 
that (for undoped vortices) the system is actually 2D. 
A complex wave function $\Psi( \mathbf{r},t)$ 
represents the superfluid helium state, with atomic
density $\rho (\mathbf{r},t)= |\Psi( \mathbf{r},t)|^2$.
To investigate the emergence of vortex structures in
this system, we look for
solutions of the time-dependent density functional equation in a
rotating frame-of-reference with constant
angular velocity $\omega $ (co-rotationg frame):

\beq
i\hbar {\partial \Psi ({\bf r},t)\over \partial t}=
[\hat {H}-\omega \hat{L}_z]\Psi  ({\bf r},t)
\label{eq1}
\eeq
where $\hat{L}_z$ is the $z$-component of the orbital angular momentum
operator. In the above equation, $\hat {H}$ is the 
DFT Hamiltonian resulting from the
functional variation of the energy density functional of 
Ref. \onlinecite{Dal95}, modified  to handle
highly inhomogeneous helium density profiles 
as those appearing {\it e.g.} around 
very attractive impurities.\cite{Anc05}  More specifically,

\beq
\hat {H} = 
  -\frac{\hbar^2}{2M}\nabla^2 +
  \frac{\delta {\cal E}[\rho]}{\delta \rho(\mathbf{r})}
  \label{eq2}
\eeq
where $\cal{E} [\rho]$ 
is the potential energy density per unit volume.\cite{Dal95,Anc05}  
We have not included the velocity-dependent 
backflow term because it is ill-behaved at low densities.
Although this is expected to impact the 
dynamics of vortex creation/evolution, 
this approximation cannot affect much
the results in the present case 
where the stationary states and their relative
energies are considered.
 
Rather than nucleating vortices in the cylinder by letting
it rotate in real-time according to Eq. (\ref{eq1}),  
we follow a more efficient strategy  
looking for stationary solutions in the co-rotating frame, 
$\Psi (\mathbf{r},t) = e^{-\imath \mu t / \hbar} \Psi_0 (\mathbf{r})$,
where  the chemical potential $\mu$ and the time-independent 
effective helium wave function $\Psi_0$ 
are obtained by solving the time-independent version 
of Eq. (\ref{eq1})

\beq
[\hat {H}-\omega \hat{L}_z] \,  \Psi_0  (\mathbf{r})  =  \,  \mu \,  \Psi_0  (\mathbf{r}) 
\label{eq3}
\eeq
To determine  $\Psi_0  (\mathbf{r})$  
describing a configuration where
$N_v$ vortex lines are present
we follow  the ``imprinting''  strategy,
{\it i.e.} we start the imaginary-time evolution of Eq. (\ref{eq3})  
leading to the minimum energy configuration with a helium wave function\cite{Pi07} 

\beq
\Psi_0(\mathbf{r})=\sqrt{\rho_0(\mathbf{r})}\, \sum _{j=1}^{N_v} \left[ {(x-x_j)+i (y-y_j)
\over \sqrt{(x-x_j)^2+(y-y_j)^2}}  \right] 
\label{eq4}
\eeq
where  $\rho_0(\mathbf{r})$ is the density of the vortex-free configuration 
and $(x_j, y_j)$ is the initial position of the $j$-vortex core with 
respect to the axis of the cylinder.
We remark that during the functional minimization
the vortex coordinates $(x_j, y_j)$ will change
to provide, at convergence, the lowest energy 
vortex configuration;
we also define $r_j \equiv  \sqrt{x_j^2 + y_j^2}\,$
as the radial position of the $j$-th vortex.
Details on how Eq. (\ref{eq3}) has been solved 
can be found in Ref. \onlinecite{Her07}. 
Both the density and wave function have been 
discretized in cartesian coordinates using
a spatial grid fine enough to guarantee
well converged results. The spatial
derivatives have been calculated with 13-point formulas.
Fast-Fourier techniques\cite{Fri05} have been employed to efficiently
calculate the energy density and mean-field 
potential.

\begin{figure}[t]
\includegraphics[width=1.2\linewidth,clip=true]{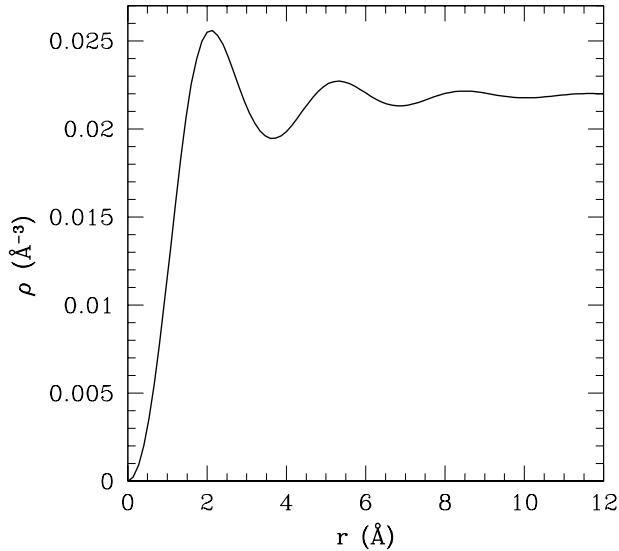}
\caption{\label{fig1} 
Vortex core structure.
}
\end{figure}

Within DFT approximation the vorticity field has a singularity
along one or several lines, the vortex cores, where the density 
vanishes and the velocity diverges. 
The helium density around one such vortex line is shown in  Fig. \ref{fig1}. 
In accordance with previous studies, the vortex structure is 
characterized by a core region of size $a_c \sim 1$ \AA{}.

\section{Results}

\subsection{Undoped vortices}

By using the imprinting method described above, for a given 
angular velocity $\omega$ we have computed a number of 
lowest energy configurations with a
fixed number of vortices $N_v$. If more than
one configuration is obtained with the same 
$N_v$ --depending on the initial guess for the 
vortex distribution embodied in Eq. (\ref {eq4})-- 
we choose the one with lower total energy.

The configurations of a vortex array in a rotating 
cylinder can be completely characterized,\cite{Hes67} 
within the Onsager-Feynman model,
by the dimensionless free energy per unit length 
${\cal F}\equiv (M/\rho \pi \hbar ^2)F$ (that at zero temperature coincides 
with the energy per unit length),
the dimensionless angular velocity 
$\Omega \equiv R^2M \omega / \hbar$ and the
scaled radial positions of the vortices $r_i/R$, 
where $\rho =0.0218$ \AA$^{-3}$
is the atom density of liquid $^4$He and $R$ the cylinder radius.
We will use these units in the following,
thus making easier to compare our results,
obtained for a system of nanoscopic size, with 
experimental results characterized
by much larger values of $R$ and much smaller values of $\omega$.

\begin{figure}[t]
\includegraphics[width=1.0\linewidth,clip=true]{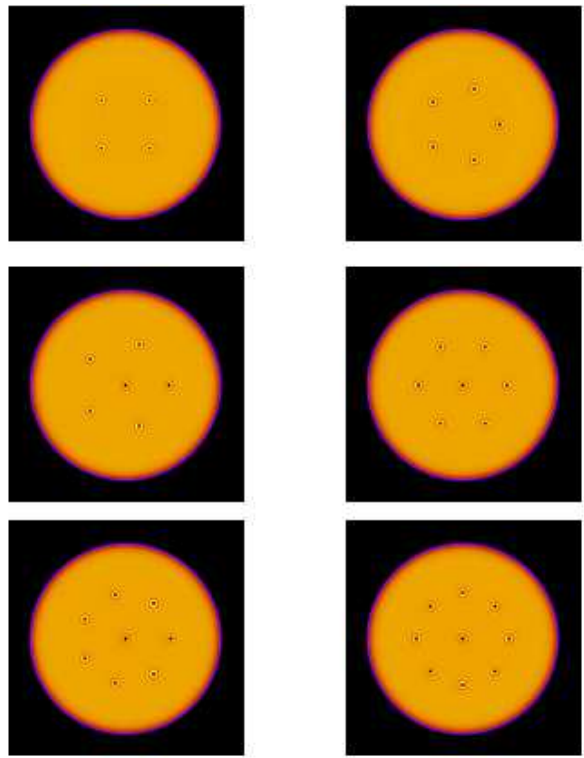}
\caption{\label{fig2} 
Stationary, lowest energy vortex configurations with  $N_v=4$ to $N_v=9$.
The portion of the simulation cell shown is $180\times 180$ \AA{} 
wide.
The color scale used to display the density values 
is the same as in Fig. \ref{fig5}.
}
\end{figure}

We show in Fig. \ref {fig2} a few stationary configurations
with $N_v = 4$ to $N_v = 9$.
The radius of the nanocylinder at rest is $R=71.4$ \AA{}. 
It has been chosen rather arbitrarily as a compromise between 
numerical affordability and
the need of disposing of a ``nanobucket'' that could host  
many vortex lines. 

\begin{figure}[t]
\includegraphics[width=1.2\linewidth,clip=true]{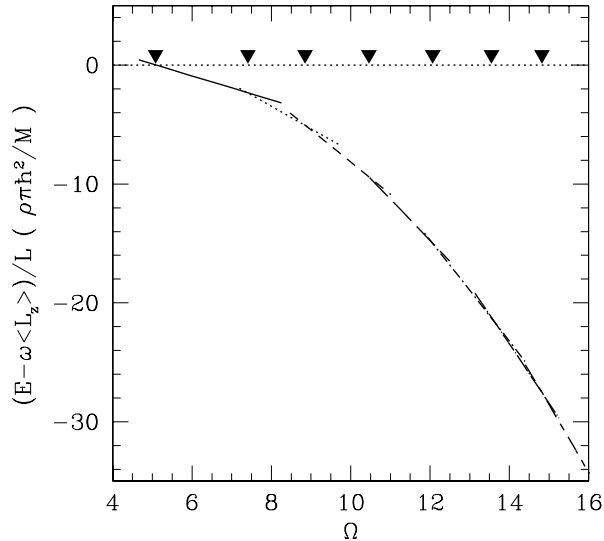}
\caption{\label{fig3} 
Stability diagram for a number of vortex lines $N_v = 0,1,2, \ldots, 7$
as a function of the dimensionless angular velocity  $\Omega = R^2M \omega / \hbar$.
The horizontal line marks the energy of the vortex-free system. 
The vertical axis is the energy per unit vortex length in the rotating frame 
expressed  in units of  $\rho \pi \hbar ^2/M$, see text.
The upside down triangles mark the crossings between different stability lines.
}
\end{figure}

The stability diagram is shown in Fig. \ref {fig3}, where
the energy per unit vortex length in the co-rotating
frame, {\it i.e.}  
$ {\cal E} \equiv (\langle \hat{H} \rangle - \omega \langle \hat{L}_z \rangle )/L$,
is shown as a function of the dimensionless angular velocity 
$\Omega $ for up to $N_v = 7$.
The crosses between the different $N_v$ lines, indicated by upside down
triangles, yield the critical rotational velocities for 
$N_v$-vortex nucleation. 

Within each stability region, the calculated energies show 
an almost linear behavior with
$\Omega$. This behavior is strictly linear for $N_v$=1, as the centered 
single vortex state is an eigenstate
of the total angular momentum. For other $N_v$ values this is 
not so and ${\cal E}(\Omega)$  bends. However,
this happens outside the corresponding stability region. As a consequence, the average  
slope of each stability region in Fig. \ref {fig3} changed of sign does represent
the total angular momentum per unit length.
DFT yields for $\langle \hat{L}_z /L \rangle/ \rho_L$, being $\rho_L$ the number of
helium atoms in the cylinder per unit length,  
a value of 1 for $N_v=1$, and of 7.081 for $N_v=9$.
It is possible to use linear response theory around 
the equilibrium configurations corresponding to
each $\Omega$ value to determine the moment of inertia  
per unit length around the symmetry axis of the 
cylinder $I_z$.\cite{Zen14,Lip03}
However, the mentioned linear behavior of 
$ {\cal E} \equiv (\langle \hat{H} \rangle - \omega \langle \hat{L}_z \rangle )/L$ 
allows one 
to obtain $I_z$ in a much simpler way, writing within each  
stability region $\langle \hat{L}_z /L \rangle/ \rho_L = I_z \Omega$.
$I_z$ displays a step-like behavior as a function of $\Omega$, 
being zero in absence of vortices.

It is illustrative to compare the DFT results with those 
obtained using the classical vortex theory of inviscid and 
incompressible fluids.\cite{Hes67} 
It turns out that both yield results in 
agreement with each other. In particular, 
the DFT values for the total angular momentum per unit length
expressed in reduced units, 
${\cal L}=\langle \hat{L_z}\rangle/(\rho \pi \hbar R^2)$, are
very close to the classical theory ones given by the expression\cite{Hes67} 
$ \sum _{i=1}^{N_v}(1-r_i^2/R^2)$, where $r_i$ is 
the distance of the $i$-th vortex from the rotation axis. 
Indeed, for the values of $N_v$ shown in Fig. \ref {fig3}
they agree to within $\lesssim 1\%$.

However, a larger discrepancy is found for the critical 
rotation velocity for the nucleation of
a single vortex that  within the 
classical vortex theory is given by\cite{Don91}

\beq
\omega_c = \frac{\hbar}{MR^2} \, ln \left(\frac{R}{a_c}\right)
\label{eq5}
\eeq
hence $\Omega_c = ln (R/a_c)$. 
Using our system values for 
$R$ and $a_c$ the above equation yields $\Omega_c=4.3$, whereas
the DFT value, given by
the intersection of the $N_v=1$ line in Fig. \ref{fig3}
with the horizontal line representing the vortex-free energy, is $\Omega_c=5.1$.
There would be needed 
much smaller a vortex core value ($a_c = 0.44$ \AA{}) to 
reconcile the classical theory with the DFT results. 
 
In the case of two linear vortices
symmetrically placed with respect to the axis of the 
cylinder, the energy of the pair as computed from the 
classical theory is\cite{Hes67}

\beq
E_2= {2\rho \pi \hbar ^2\over M}\,
 \left[ln\left(\frac{R}{a_c}\right)+ ln\, 
\frac{(1-p^2)}{2}-{1\over 2} ln\, p-\Omega (1-p)\right]
 \label{eq6}
 \eeq
where $p\equiv (d/2R)^2$ with $d$ being the vortex-vortex distance. 
The equilibrium condition $dE_2/dp=0$ yields

\beq
3p^2+1-2\Omega p(1-p^2)=0
\label{eq7}
\eeq
which admits a $p>0$ solution 
as long as $\Omega > \Omega _0=\sqrt{9/4+3\sqrt{3}/2}=2.202$.

\begin{figure}[t]
\includegraphics[width=1.2\linewidth,clip=true]{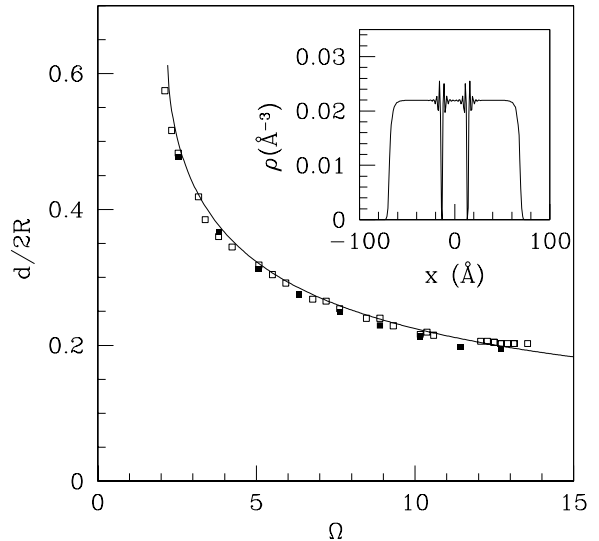}
\caption{\label{fig4} 
Vortex-vortex equilibrium distance for the two-vortex array 
as a function of the dimensionless angular velocity  $\Omega = R^2M \omega / \hbar$.
Open squares: DFT result for the empty vortex cores. 
Filled squares: DFT result for the Xe-filled vortex cores.
Solid line: vortex model result, Eq. (\ref{eq7}).
The inset shows the DFT 
empty vortex-pair equilibrium configuration at the closest approach.
}
\end{figure}

We plot in Fig. \ref{fig4} the calculated vortex-vortex  
equilibrium distance for the two-vortex array
as a function of the angular velocity and compare it 
with the value obtained from Eq. (\ref{eq7}).
The agreement is good at low $\Omega $, up to the limiting value
$\Omega _0$ (which is the lowest displayed value 
with the solid line).
At high values of $\Omega $ the DFT results level off.
This is due to the superposition of the core structures
which prevents further decrease in the distance, and the classical vortex
theory does not hold because of the inhomogeneities 
in the density profile, shown in the inset of Fig. \ref{fig4},
where the vortex structure at closest approach is displayed.
For angular velocities larger than those displayed in the figure
with open squares, the whole
$^4$He cylinder becomes unstable and the DFT minimization
procedure fails. 

As for the equilibrium structures, the DFT results
are in agreement with those of 
the classical vortex theory\cite{Cam79}
for a rotating cylinder of superfluid $^4$He.
The energetically
favored structures for $N_v> 5$ are made of rings of vortices 
plus one vortex at the center of the cylinder.
The tendency of rings of vortices to form was observed long
ago in a rotating bucket experiment.\cite{Yar79}
Our findings are consistent with this observation and with other fine details. 
In particular, for $N_v=6$ besides the stable five-fold ring 
of vortices plus a vortex at the center,  
a metastable state made of a six vortex ring is experimentally observed; in our 
calculations this state  is almost
degenerate with the stable one. Both configurations were 
also found within the classical vortex theory,\cite{Cam79} 
but with the six-vortex ring fairly higher in energy than 
the stable state.

\begin{figure}[t]
\includegraphics[width=1.1\linewidth,clip=true]{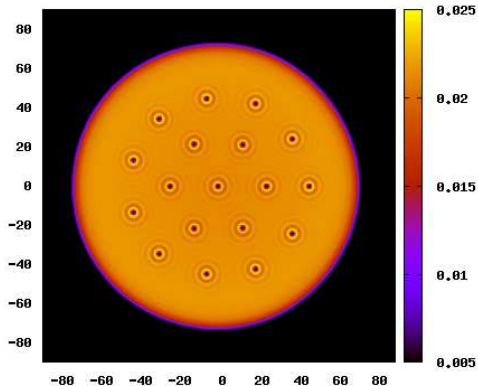}
\caption{\label{fig5} 
 $N_v = 18$  state at $\Omega =29.6$.
Lengths are in \AA{}.
The vertical scale shows the displayed values of density,
between $\rho =0$ and $\rho =0.03$ \AA $^{-3}$.
}
\end{figure}

A configuration with a larger number of 
vortices, namely $N_v = 18$, is shown in
Fig. \ref{fig5}. This equilibrium vortex structure 
again coincides with the lowest energy structure 
of classical vortex theory.\cite{Cam78} 
Within such theory, the areal density of vortex 
lines $n_0$  is proportional to the angular velocity,  
$n_0=2M \omega /h=\Omega /\pi R^2$.\cite{Fey55,Sta68} 
Assuming a triangular distribution for the vortex lines,
the areal density would be $n_0=2/(\sqrt{3}d^2)$, 
where $d$ is the mean inter-vortex distance.
By equating these two expression for $n_0$, with 
the value $\Omega = 29.6$ used to obtain the 
distribution shown in Fig. \ref{fig5} one gets
$d/R=\sqrt{2\pi/\sqrt{3}\Omega}=0.35$.
From Fig. \ref{fig5} one can estimate an
average vortex-vortex distance $d\sim 24$ \AA{},  {\it i.e.}
$d/R=0.34$, which compares well with the 
result of the classical vortex theory. 

We remark at this point that the scaled lengths and
frequencies $r/R$ and $\Omega \equiv R^2M \omega / \hbar$ 
which characterize the vortex array configurations\cite{Hes67}
allow to compare the results for a nanoscopic
system, like the ones presented here, to the 
actual experiments where typical lengths and 
frequencies differ by many orders of magnitude.
This is proven, for instance, by looking at the rotating bucket
experimental results of Ref. \onlinecite{Yar79}.
Figure 2(e) in this reference
shows the 5-fold ring of vortices nucleated 
in a rotating bucket of radius $R=1$ mm with
angular velocity $\omega = 0.45\,s^{-1}$. The average
scaled distance between neighboring vortex cores 
can be read directly from that figure,
$d/R\sim 0.32-0.33$.
From our DFT calculations for a 5-fold ring 
of vortices at the same value of the 
dimensionless frequency $\Omega =28.5$
in a nanobucket with $R=71.4$ \AA{} we get
a ratio  $d/R=0.34$ which is compatible with 
the experimental one.

\subsection{Doped vortices}

We next study the changes in the
vortex structures induced by the capture of 
atomic impurities inside the vortex cores.
We consider the particular case of Xe atoms because 
of their use as vortex tracers in recent experiments.\cite{Gom14} 
Due to the large mass of the Xe atom as compared 
to that of the He atom, their effect 
on the liquid is incorporated through an external potential 
$V_{\rm He-Xe}$ (which is taken from Ref. \onlinecite{Tan03}), {\it i.e.}
$\hat{H}$ in Eq. (\ref{eq2}) is replaced by  
$\hat{H} + \sum_I V_{\rm He-Xe}(|\mathbf{r} - \mathbf{R}_{\rm I}|)$,
where $\mathbf{R}_{\rm I}$ is the position of the $I$-th Xe atom.

\begin{figure}[t]
\includegraphics[width=1.1\linewidth,clip=true]{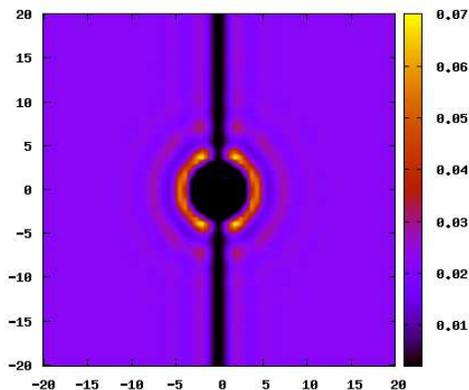}
\caption{\label{fig6}
Helium density around a Xe impurity trapped inside a vortex line.
Lengths are in \AA{} and densities in \AA$^{-3}${}.
}
\end{figure}

The equilibrium density profile around a Xe impurity 
trapped inside a vortex core is shown in Fig. \ref{fig6}. 
Due to the periodic boundary conditions inherent to the  
use of the Fast-Fourier method,\cite{Fri05}
this configuration actually corresponds
to a linear chain of Xe atoms separated 
one-another by a distance 
equal to the length of the simulation cell along
the vortex axis, which in the present case is 30 \AA{}. 
Since the Xe distance between 
periodically repeated images is so large, the interaction
between images can be safely neglected and in practice that configuration		
representis indeed an {\it isolated} Xe atom attached to the vortex.
We have calculated the binding energy
of the Xe atom to the vortex line as\cite{Pi07}

\beq
\label{eq8}
B_{\rm Xe}=(E_{\rm Xe} - E_0) - ( E_{{\rm Xe}+V}-E_V)
\eeq
where
$E_{{\rm Xe}+V}$, $E_{\rm Xe}$, $E_V$, and $E_0$
are the energies of the (vortex+Xe), (Xe), (vortex) and
pure $^4$He cylinder, respectively.
We have found $B_{\rm Xe}=3.2$ K, which compares 
favorably with earlier estimates,\cite{Dal00} 
where a value close to $B_{\rm Xe} = 5$ K was found 
using a different functional.
The positive value of $B_{\rm Xe}$ implies that the Xe impurity is  
energetically stabilized inside the vortex line.

\begin{figure}[t]
\includegraphics[width=1.2\linewidth,clip=true]{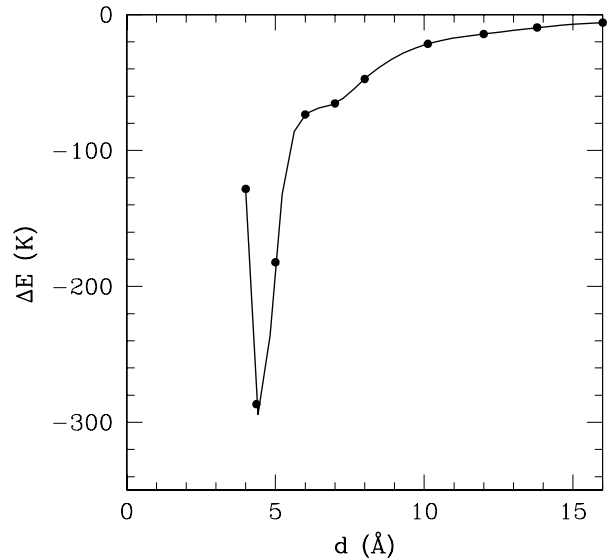}
\caption{\label{fig7} 
Energy  of two xenon atoms in a vortex line {\it vs.} Xe-Xe distance $d$.
}
\end{figure}

\begin{figure}[t]
\includegraphics[width=1.0\linewidth,clip=true]{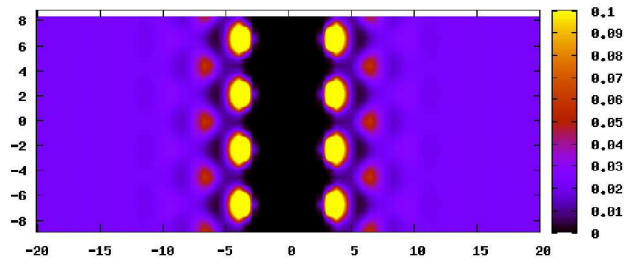}
\caption{\label{fig8}
Xe chain embedded inside a vortex line.
The chain is made by a periodically repeated motif
of four equidistant Xe atoms at $r=0$ and  $z= -8.8, -4.4, 0$, and 4.4  \AA.
The mutual Xe-Xe distance is chosen as the equilibrium one for 
the Xe pair inside a single vortex line.
Lengths are in \AA{} and densities in \AA$^{-3}${}.
}
\end{figure}

We have also computed the energy of two Xe atoms within the
same vortex line as a function of the Xe-Xe distance. 
To model the Xe-Xe interaction we have used
the pair potential function computed in Ref. \onlinecite{Tan03}.
The results are shown in Fig. \ref{fig7}, where the 
energy difference with respect to the configuration of two 
Xe atoms well apart from each other is shown as a 
function of the atoms separation.
As was also found for other impurities,
\cite{Pop13} the Xe atoms are free to move
along the vortex line, and the lowest energy state
is the one where the Xe atoms have formed
a dimer, whose bond length almost coincide
with that of the Xe dimer in vacuum.

Additional Xe atoms trapped within the same
vortex line can in principle form a 
one-dimensional atomic chain completely filling the
vortex core. The structure of such Xe chain 
is shown in Fig. \ref{fig8}. The helium density close to the
Xe atoms appear to be rather structured, with values
that locally exceed the bulk density by a factor of 2-3.
It was experimentally found that the atomic
impurities are trapped in  
vortex lines in the form of regularly spaced atomic clusters, 
rather than forming atomic chains.\cite{Gom12,Spe14,Lat14,Tha14} 
Actually, the relatively large size of such clusters
allow to use them to effectively image the vortex 
itself.\cite{Gom14}

While being certainly interesting because of its relevance 
to the experimental studies of the elusive 
vortices in superfluid $^4$He nanodroplets,
the theoretical study of atomic clusters in vortex lines
is beyond the scope of this paper.
We rather address briefly here the simpler case of a 
vortex line filled with Xe atomic chain.
Albeit being aware of its limitations,
we believe nevertheless that it might be a useful  
first attempt to address the rather complex issue
of impurity aggregates inside vortex arrays. 
It is worth mentioning that cluster merging 
inside the same vortex line may be hampered
by the existence of energy barriers --to which the 
very structured helium density around impurities 
contribute in a non-negligible way--
and that there are experimental\cite{Prz08,Log11} and 
theoretical\cite{Elo08,Her08} examples of 
metastable structures made of nearly isolated impurities 
or impurity clusters coexisting in helium droplets. 
The specific characteristics of the 
formation of atomic clusters in helium droplets haven 
been reviewed in Ref. \onlinecite{Tig07}.

\begin{figure}[t]
\includegraphics[width=1.1\linewidth,clip=true]{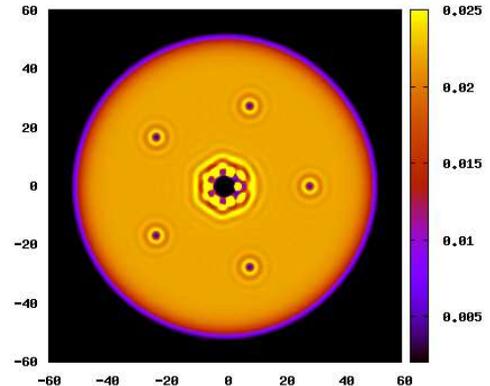}
\caption{\label{fig9}
Vortex structure around an annulus made  
by filling  the central vortex line with Xe atoms at the
dimer distance. The angular velocity is $\Omega= 19$
and the radius of the rotating $^4$He cylinder is $50\,\AA$.
Lengths are in \AA{} and densities in \AA$^{-3}${}.
}
\end{figure}

By completely filling the core of a single vortex  
with a chain of Xe atoms at the dimer equilibrium 
distance, the liquid helium is expelled from 
the region around the axis of the cylinder constituting
an annulus of inner radius about that of the Xe-He 
pair-potential core that replaces the vortex line.
An annular geometry was used by Vinen in his classical experiment on 
quantized circulation,\cite{Vin61} showing that above a 
certain angular velocity a quantized 
circulation of the superfluid velocity appeared 
around the axis of the annulus, and that increasing further 
the angular velocity vortices could appear.  
Low-lying states of rotating superfluid $^4$He in an annulus were 
studied by  Stauffer and Fetter\cite{Sta68}
with the classical inviscid fluid model finding  
that the vortices lie on a ring midway between the 
boundaries of the annulus. The number of vortices in the 
ring increases with increasing angular velocity, with the 
possibility of forming more than one ring.
The same phenomenology appears in DFT simulations, 
as shown in Fig. \ref{fig9}, where we show the calculated 
structure with a 5-vortex ring enclosing the annulus
in the center formed by a Xe-filled central vortex line.

The filling of neighboring vortex lines with atomic
impurities/clusters is likely having observable
effects on the vortex distribution in a multi-vortex
configuration, like the ones recently observed 
in $^4$He nanodroplets. \cite{Gom14} 
As a first step towards understanding the effect of
cluster doping on a vortex array, we consider here the 
interaction between a pair of doped vortices,
similarly to what done for the ``empty'' vortex pair,
but with the cores completely filled by
a chain of Xe atoms.
Any effect should show up in changes of 
the vortex mutual distance as a function of the rotational
frequency, as compared to the case of empty vortices.
The results of our calculations are shown in Fig. \ref{fig4} with 
filled squares.
It appears that there is no evident change in the 
vortex-vortex distance induced by the
Xe adsorption (apart when the two vortex cores are
very close to one another), in spite of the additional rotational 
energy $M_L(d/2)^2\omega ^2$ 
due to the Xe mass ($M_L$ being the Xe mass per unit vortex length)
rotating with the vortex pair.
Although such contribution is small in the present case, due to the
nanoscopic dimension of our system, it could become
relevant in experimental situation, altering the distribution
of vortex lines containing Xe clusters, especially at the
periphery of the droplet. Such effect seems to have been observed 
in the experimental images of Ref. \onlinecite{Gom14}.

\section{Summary and outlook}

Within the zero temperature Density Functional approach, we have studied
the formation of vortex arrays
in a rotating $^4$He cylinder of nanoscopic dimension.
We have found that the simple scaling relations that characterize the classical 
theory of quantized vortices in incompressible and
inviscid fluid can be used to determine, starting from the nanoscale
DFT results presented here, the structure of vortex arrays in the
millimeter-sized samples used in rotating bucket experiments.

Motivated by current experiments on $^4$He nanodroplets, 
we have also addressed the effect of doping the vortex 
cores with Xe impurities. Somewhat unexpectedly, we have 
found that adding these impurities does not introduce sensible changes
in the inter-vortex distance. Since such changes has  
been experimentally observed at the
periphery of droplets,\cite{Gom14} there 
remains to be seen whether they are due to the role played by the 
geometry: unlike the case of an infinitely extended 
cylinder, in a spherical drop quite some Xe atoms/clusters
are located not far from its curved surface where the 		
vortex cores are wider and the helium density lower.
We are currently undertaking the study of vortex 
arrays within $^4$He nanodroplets,
which will be the subject of a forthcoming work.

\acknowledgments
We thank Andrei Vilesov, Jes\'us Navarro and Andrew Ellis
for stimulating discussions.
This work has been performed under Grants No. FIS2011-28617-C02-01
from DGI, Spain (FEDER) and  2014SGR401 from Generalitat de Catalunya.

\end{document}